\documentclass[useAMS,usenatbib]{mn2e}
\usepackage{graphicx}

\newcommand{\ms}{$\,$M$_\mathrm{\odot}$}
\newcommand{\ls}{$\,$L$_\mathrm{\odot}$}
\newcommand{\be}{\begin{equation}}
\newcommand{\ee}{\end{equation}}
\newcommand{\delrad}{\bigtriangledown_\mathrm{rad}}
\newcommand{\dela}{\bigtriangledown_\mathrm{ad}}

\title[Deep Dredge-up in Intermediate-Mass TP-AGB Stars]{Deep Dredge-up
in Intermediate-Mass Thermally Pulsing AGB Stars}
\author[R.J. Stancliffe, C.A. Tout and O.R. Pols]{Richard J. Stancliffe\thanks{E-mail:
rs@ast.cam.ac.uk}$^1$, Christopher A. Tout$^1$ and Onno R. Pols$^2$\\
$^1$Institute of Astronomy, The Observatories, Madingley Road, Cambridge CB3 0HA\\
$^2$Sterrekundig Instituut Utrecht, Postbus 80000, 3508 TA Utrecht, The Netherlands}
\begin{document}

\date{Accepted 0000 December 00. Received 0000 December 00; in original form 0000 October 00}

\pagerange{\pageref{firstpage}--\pageref{lastpage}} \pubyear{0000}

\maketitle

\label{firstpage}

\begin{abstract}
We present results of the evolution of AGB stars of 3\ms\ and 5\ms\ with solar metallicity calculated with the Eggleton stellar evolution code ({\sc{stars}}), which has a fully implicit and simultaneous method for solving for the stellar structure, convective mixing and nuclear burning. We introduce the concept of a viscous mesh in order to improve the numerical stability of the calculations. For the 5\ms\ star, we evolve through 25 thermal pulses and their associated third dredge-up events. We obtain a maximum helium luminosity of $1.7\times10^9$\ls\ and significantly deep dredge-up after the second pulse. Strong hot-bottom burning is observed after the $5^{\mathrm{th}}$ pulse. The 3\ms\ model is evolved through 20 thermal pulse events and we find third dredge-up after the 7th pulse. During the 14th pulse sufficient carbon has been brought to the surface to produce a carbon star. We find that dredge-up and the transformation into a carbon star occur at significantly smaller core masses (0.584\ms\ and 0.608\ms, respectively) than in previous calculations for 3\ms.
\end{abstract}


\section{Introduction}

The observed enhancements of carbon and $s$-process elements in asymptotic
giant branch (AGB) stars show that material, processed by helium burning
during thermal pulses, is being dredged up to the surface.  Although
qualitatively well understood, the extent and efficiency of this third
dredge-up (TDUP) process has been very hard to model quantitatively, and
remains one of the major uncertainties in AGB evolution. The amount of TDUP obtained for a star of a certain mass and composition differs greatly between one model calculation and another. It depends on many factors, such as the numerical treatment of convective boundaries \citep{Lat89}, whether allowance for extra mixing is made \citep{Hol88}, the mixing-length parameter \citep{BSIV88} and the numerical resolution of the mass grid and time step \citep{Stran97}.  Recently, very efficient dredge-up for stars with small core masses has been found by various authors (Herwig et al.\ 1997, Mowlavi 1999, Herwig 2000) if a diffusive form of convective overshooting is taken into account.

In the calculation of the evolution of stars of $1-8$\ms\ through the thermally pulsing asymptotic giant branch (TP-AGB) phase of their evolution most evolutionary codes use only a partially self-consistent approach. The treatment of structure, mixing and nucleosynthesis is not simultaneous, with mixing often being solved in a separate iteration step from the other two. Examples include the codes of \citet{Stran97}, \citet{WG98}, \citet{Her00} and \citet{Amand02}. As the phenomenon of third dredge-up depends critically on the treatment of convection within a stellar structure calculation it is desirable to combine the calculation of nucleosynthesis, mixing and structure into a single, simultaneous step. The first attempt to do this was made by \citet{Pols01}. In their calculations a 5\ms\ star was evolved without mass loss through the first 6 thermal pulses and deep dredge-up occured.

The occurence of deep dredge-up is important for the formation of carbon stars. These are defined as stars that are M-type and have surface carbon-to-oxygen abundance ratios exceeding unity. Observational evidence suggests that these stars are of low mass, most likely between 1 and 3\ms\ (Iben 1981). However, there has been considerable difficulty in producing detailed theoretical models of carbon stars with low enough masses and luminosities. Of the early work on the subject, \citet{BSIV88} were able to produce two carbon star models from initial masses of 1.2 and 2.0\ms\ under metal-poor conditions ($Z=0.001$), while \citet{Lat89} produced a model initially of $M=1.5$\ms\ with $Z=0.02$. More recently low-mass carbon star models have been produced by \citet{Stran97} and, with the aid of convective overshooting, \citet{Her00}. However, the core masses found in these models are still too large to explain the carbon star luminosity function (Izzard \& Tout 2004). 

While the approach of Pols \& Tout enabled them to make significant progress with a fully implicit and simultaneous treatment of TP-AGB calculations, problems with numerical stability prevented them from evolving their model further than a few thermal pulses. In this work we have developed methods that overcome some of the numerical instability that occurs in a fully implicit treatment of the evolution. These methods are described in section 2. Results of the calculations for a 5\ms\ star are presented in section 3 and compared to other available models in section 4. In section 5 we present results of the calculations for a 3\ms\ and detailed comparisons are made to other models in section 6.

\section{The Stellar Evolution Code}
The models were calculated using the {\sc{stars}} 1D stellar evolution code originally developed by \citet{PPE71a} and updated by many contributors (e.g. Pols et. al. 1995). This code solves the structural and composition equations, including the effects of both mixing and burning, implicitly and {\it simultaneously}. We compute the structure of a star over a given number of mesh points which are placed in the most appropriate regions determined by equal intervals of a mesh spacing function $Q(P,T,r,m)$ where $P,\:T,\:r$ and $m$ are pressure, temperature, radius and mass respectively \citep{PPE71a}. This function is designed to push more mesh points into regions where they are needed such as burning shells and ionization zones. The mesh is adaptive so that as the star evolves, the mesh points move around to stay with physically important regions. We apply a mesh-spacing function
\begin{eqnarray}
Q  = & \ln \left( \frac{(m/M)^{2/3}}{c_1} + 1 \right)
  \; + \; c_2 \ln \left( \frac{r^2}{c_3} + 1 \right)
  \; \nonumber \\ 
& - \; c_4 \ln \frac{T}{T + c_5} 
 - \;\; c_6 \ln P
  \; - \; c_7 \ln \frac{P + 0.1 P_\mathrm{H}}{P + 3 P_\mathrm{He}}
  \; \nonumber \\ 
 & - \; c_8 \ln \frac{P + 0.3 P_\mathrm{He}}{P + 3 P_\mathrm{He}}
\end{eqnarray}
where $c_1\ldots c_8$ are appropriately chosen constants, $M$ is the total 
mass and $P_\mathrm{H}$ and $P_\mathrm{He}$ are the values of $P$ at the 
position of the H- and He-exhausted core mass, respectively. 
The last two terms are included only during the TP-AGB phase. They concentrate 
a large part of the mesh, about three-fifths, in the intershell region 
(between $0.1P_\mathrm{H}$ and $3 P_\mathrm{He}$) while at the same time 
ensuring that during a TP, when $T$ and $P$ undergo rapid changes, 
the mesh does not move around too much in the intershell region and so that numerical diffusion is suppressed. The mesh-spacing equation is also solved implicitly and simultaneously with the structure and composition equations.

The code includes mixing and nuclear burning using a diffusion equation approximated by an implicit second-order difference equation of the form
\begin{eqnarray}
\sigma_{k+\frac{1}{2}}\frac{X_{k+1}-X_k}{\delta m_{k+\frac{1}{2}}} -
\sigma_{k-\frac{1}{2}}\frac{X_k-X_{k-1}}{\delta m_{k-\frac{1}{2}}} \nonumber \\
 = \left( \frac{X_k -
X_k^0}{\Delta t} + R_{X,k} \right)\delta m_k,
\end{eqnarray}
where $X_k$ and $X_k^0$ are the abundances at meshpoint $k$ at the present
and previous timestep, $\Delta t$ is the timestep, $\delta m_{k\pm\frac{1}{2}}$ are the masses contained in the zones above and below meshpoint $k$, 
$R_{X,k}$ is the net consumption rate of $X$ by nuclear reactions, and
$\sigma_{k\pm\frac{1}{2}}$ are the diffusion coefficients corresponding to these zones. The linear diffusion coefficient $D$ is related to $\sigma$ by 
$\sigma = (4\pi r^2\rho)^2 D$, where $r$ and $\rho$ are radius and density.

Convective mixing is is treated in the framework of the mixing-length theory 
\citep{BV}, which gives $D=D_\mathrm{MLT}=\frac{1}{3}vl$ with $v$ and $l$ the 
mean velocity and mean free path of convective eddies.
In order to maintain numerical stability, \citet{Pols01} found it useful to 
express $D$ as 
\be
D = D_\mathrm{MLT}{\beta W\over{1-(1-\beta)W}}
\ee
with
\[
W \equiv {\delrad - \dela \over{\delrad}},
\]
The value of $\beta$ was treated as a free parameter and allowed to take on different values in the H-rich envelope and in the H-free intershell region and core.

All the quantities including $\sigma$ are defined only at each meshpoint $k$ so it is not a priori obvious how $\sigma_{k\pm\frac{1}{2}}$ should be calculated. In our standard calculations we take $\sigma_{k+\frac{1}{2}} = \frac{1}{2}(\sigma_k + \sigma_{k+1})$, following \citet{PPE72}. \citet{Pols01} investigated the effect of taking a geometric mean $\sigma_{k+\frac{1}{2}} = \sqrt{\sigma_k\cdot\sigma_{k+1}}$ rather than an arithmetic mean. These schemes effectively differ only in zones where the Schwarzschild boundary is located between meshpoints $k$ and $k+1$. The arithmetic mean in effect allows the radiative meshpoint adjacent to a convective boundary to be mixed while the geometric mean does not.  The choice is usually inconsequential but becomes important in the presence of a composition discontinuity that leads to a discontinuity in $\delrad\ $through the opacity dependence.  Such discontinuities arises during core-helium burning at the edge of the convective core and at the bottom of the convective envelope during TDUP (Paczy\'nski 1977). The standard arithmetic mean ensures that $\delrad - \dela$ always approaches zero from the convective side of a boundary, even in the case of
a discontinuity, so a convective boundary always corresponds to a \emph{stable} Schwarzschild boundary ($\delrad=\dela$). However when there is a discontinuity in $\delrad$ at the boundary even the slightest extra mixing makes it unstable and physically we expect mixing to extend until $\delrad\leq\dela$ when material can mix across the boundary. Pols and Tout demonstrated that use of the geometric mean for $\sigma_{k\pm\frac{1}{2}}$ suppresses this physical behaviour and prevents third dredge-up (see also the discussion in Mowlavi 1999 and Herwig 2000) so we always employ the arithmetic mean.

The TP-AGB is a difficult phase of evolution to calculate. As the peak of a thermal pulse is reached the code must use very small timesteps, down to the order of a few minutes. This can cause problems with the stability of the code which we  outline below. After a pulse has occured it is necessary to limit the timestep to the order of weeks in order to properly resolve the process of TDUP. After this the code needs to be able to increase its timestep significantly to deal with the long quiescent interpulse period before the next thermal pulse. Sections 2.1 and 2.2 describe in detail the numerical measures we have added with these constraints in mind.

\subsection{The Viscous Mesh}
At short timesteps, such as those required to resolve the peak helium luminosity of a thermal pulse, there is a numerical instability associated with the luminosity equation
\begin{eqnarray}
 L_{k+1} - L_k = (m'E_1)_{k+\frac{1}{2}} + (m'E_2)_k[\dot{m}_k] \nonumber \\  - (m'E_2)_{k+1}[-\dot{m}_{k+1}]
\end{eqnarray}
where $L_k$ is the luminosity at mesh point number $k$. The change in mass with mesh point is denoted $m'=dm/dk$. Square brackets signify a term only included when it is positive. The first term containing $E_1$ is related to the usual energy generation terms but evaluated at constant mesh point and the terms containing $E_2$ are an upstream approximation for the advection term owing to the
adaptation of the mesh. When the timestep becomes small these terms become large and this leads to numerical instability when two large numbers are subtracted to give a small result.

To deal with this the mesh can be fixed in mass and so eliminate the last two terms of the above equation. However, it is not desirable to fix the entire mesh because the outer regions do not cause the same problems as the inner ones and the mesh need only to be fixed at small timesteps. This causes a second problem -- how do we return from the fixed to the fully adaptive mesh? The mesh cannot just be instantaneously returned to a fully adaptive state because, during the fixed phase, mesh points drift away from where the mesh spacing function would place them. Rather, the mesh must be returned to a fully adaptive state gradually and so we have developed the idea of a viscous mesh.

Our viscous mesh combines fixed and adaptive behaviours. A weighting coefficient $\gamma$, a function of mesh point number and the timestep size, is used to determine the nature of the mesh. In problematic areas the mesh is gradually fixed by increasing the value of $\gamma$. If $\gamma=1$ a mesh point is fully fixed in mass while if $\gamma=0$ the point is fully adaptive. We implement this by solving
\be
 \left(\left(\frac{dQ}{dk}\right)_{k+1}-\left(\frac{dQ}{dk}\right)_{k}\right)(1 - \gamma) + \gamma\left(\frac{dm}{dt}\right)_k = 0
\ee
alongside the equations of stellar structure and composition. We choose the weighting coefficient $\gamma$ to be a function of the timestep with $\gamma$ becoming unity for timesteps of $10^{-4}$ yrs and below for the central mesh points.

\subsection{Timestep control}
The {\sc{stars}} code attempts to choose the most appropriate timestep size based on the changes to the variables of the previous model required to produce the current model. A sum of the absolute values of these changes (excluding those made to the luminosity) is made over all variables and over all meshpoints producing a single numerical value 
\[
 d = \Sigma_i \Sigma_k |x_{i,k}|,
\]
where $x_{i,k}$ are the values of the changes made to each variable $i$ at a given meshpoint $k$. Usually the value of $d$ is dominated by the temperature and the degeneracy. This is then compared to a preset optimum value $d_\mathrm{opt}$. If $d_\mathrm{opt}/d$ is greater than one the timestep is increased by this fraction or 1.2, whichever is smaller. If $d_\mathrm{opt}/d$ is less than one the timestep is reduced by this fraction or 0.8, whichever is larger. In this way the most appropriate timestep is chosen for the next model.  

 The value chosen for $d_\mathrm{opt}$ depends on the number of mesh points in the model. For a 999 mesh point model, a value of 5 is typically chosen. However, if it is desirable to have smaller timesteps then lower values are used. In order to deal with the range of timescales involved in TP-AGB evolution we find it necessary to change the value of $d_\mathrm{opt}$ at certain stages of the evolution.

As the peak of a thermal pulse is approached the timestep drops to about $2\times10^{-5}$ years. To avoid numerical problems inherent at lower timesteps we need to boost $d_\mathrm{opt}$. We typically do this by adding $0.5$ when the timestep drops below $2.5\times10^{-5}$ years. When the helium luminosity begins to decline after the peak of the thermal pulse the intershell convection zone (ICZ) shuts down and the convective envelope of the star moves inward. During this phase $d_\mathrm{opt}$ is slowly returned to 5. As the envelope reaches down into the region where the ICZ was active it begins to dredge-up the products of helium burning. At this point it is necessary to limit the size of the changes being made to the variables to avoid numerical instability.

We have written an algorithm that records the position of the boundary of the ICZ at its maximum outward extent. We then compare this to the location of the boundary of the convective envelope. When the envelope reaches into the region where the ICZ was active the value of $d_\mathrm{opt}$ is set to 1. This limits the timestep to the order of a few days. When the helium luminosity reaches as low as $3\times10^3$\ls\ and TDUP is over $d_\mathrm{opt}$ is restored to 5.

\section{Calculations for a 5\ms Star}
We apply our code to a 5\ms model with $Z=0.02$, evolved from the pre-main sequence with 199 mesh points and solar composition. Before the onset of the TP-AGB phase we increase the number of mesh points to 999 to provide sufficient spatial resolution in the intershell region. Throughout the evolution we do not consider mass loss nor do we include convective overshooting. We use the formalism of \citet{BV} for convection with the ratio between the mixing length and the pressure scale height $\alpha=1.925$. We choose our $\beta=1$ in the envelope of the star, and $\beta=5\times10^{-5}$ in regions where the mass fraction of hydrogen is less than $10^{-6}$. As determined by \citet{Pols01} there is negligble quantitive difference if we increase these values.

\begin{figure}
\includegraphics[width=7cm, angle=270]{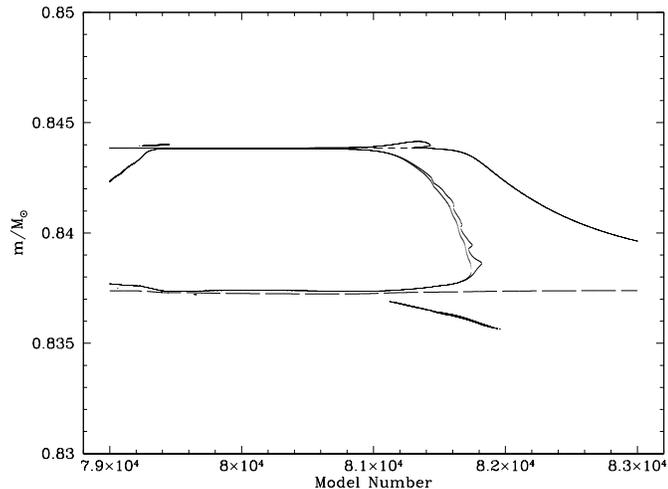}
\caption{Plot showing a spurious convection zone below the helium burning shell. The boundaries of convection zones are marked with solid lines. The hydrogen and helium burning shells are marked with short and long dashed lines respectively. The zone occurs in a region where the temperature gradient is positive and so does not represent a true convective region. The upper zones cannot be seen in this figure as they lie between the top of the ICZ and the hydrogen burning shell.}
\label{fig:spuriousconv}
\end{figure}

In the course of these evolutionary calculations problems were encountered with the composition equations. These were found to be related to the spontaneous creation of convective zones above and below the ICZ. An example of the lower zone is shown in Figure \ref{fig:spuriousconv}. The zones below the ICZ occur in regions where the temperature gradient is positive and hence are not true convective regions. At most these zones are no more than a couple of mesh points wide and their lack of extent confirms that they are not failed degenerate pulses as described in the work of \citet{Frost98}. These zones begin to occur after the 7$^{\mathrm{th}}$ thermal pulse but do not cause serious problems with model convergence until the 12$^{\mathrm{th}}$ pulse.

In order to eliminate these spurious zones we apply the following strategy. At timesteps below $4\times10^{-4}$ years we reduce the value of $\beta$ to $10^{-7}$ for the 25 mesh points immediately above the top of the ICZ. We do this because we expect this zone to be radiative and so no mixing should occur there. The spurious zones do not then appear. Below the helium burning shell we prevent the use of both convective mixing and the transport of energy by convection. These measures have been tested both on the 6 pulses where no spurious zones were found and on pulses 7 to 12 where they occurred. The evolution is the same both with and without them. Thus we are confident that they do not affect the evolutionary calculations.

\begin{figure}
\includegraphics[width=8cm]{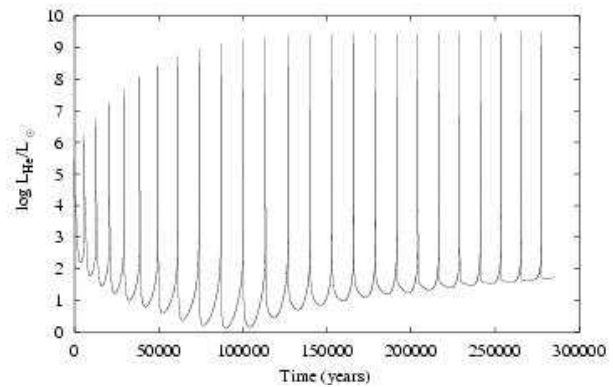}
\caption{Plot of the evolution of the helium luminosity versus time since the peak of the first thermal pulse in years, for the 5\ms\ model.}
\label{fig:Hevt}
\end{figure}

\begin{figure}
\includegraphics[width=7cm]{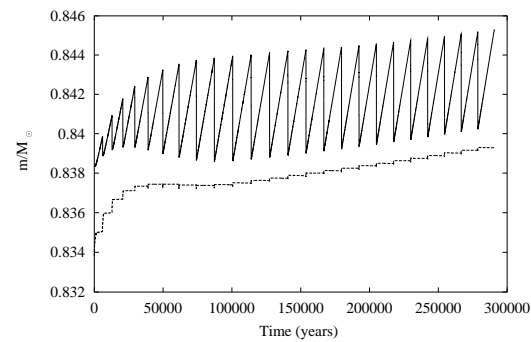}
\caption{The evolution of the hydrogen-exhausted (solid line) and helium-exhausted (dashed line) core masses with time since the peak of the first thermal pulse, for the 5\ms\ model.
}
\label{fig:5core}
\end{figure}

With the outlined modifications, a total of 25 thermal pulses and their associated TDUP episodes were calculated. A plot of the evolution of the helium luminosity with time is shown in Figure \ref{fig:Hevt} and the evolution of the hydrogen- and helium-exhausted core masses with time is shown in Figure \ref{fig:5core}. These core masses are defined as the mass coordinates where the mass fractions of H and He are less than 0.3. Thermal pulses begin when the helium core mass is 0.838\ms. We find the peak helium luminosity rapidly approaches $1.7\times10^9$\ls. Third dredge-up occurs on the very first pulse and significant hot-bottom burning is seen after the 4$^\mathrm{th}$ pulse.

\begin{figure*}
\includegraphics[width=15cm]{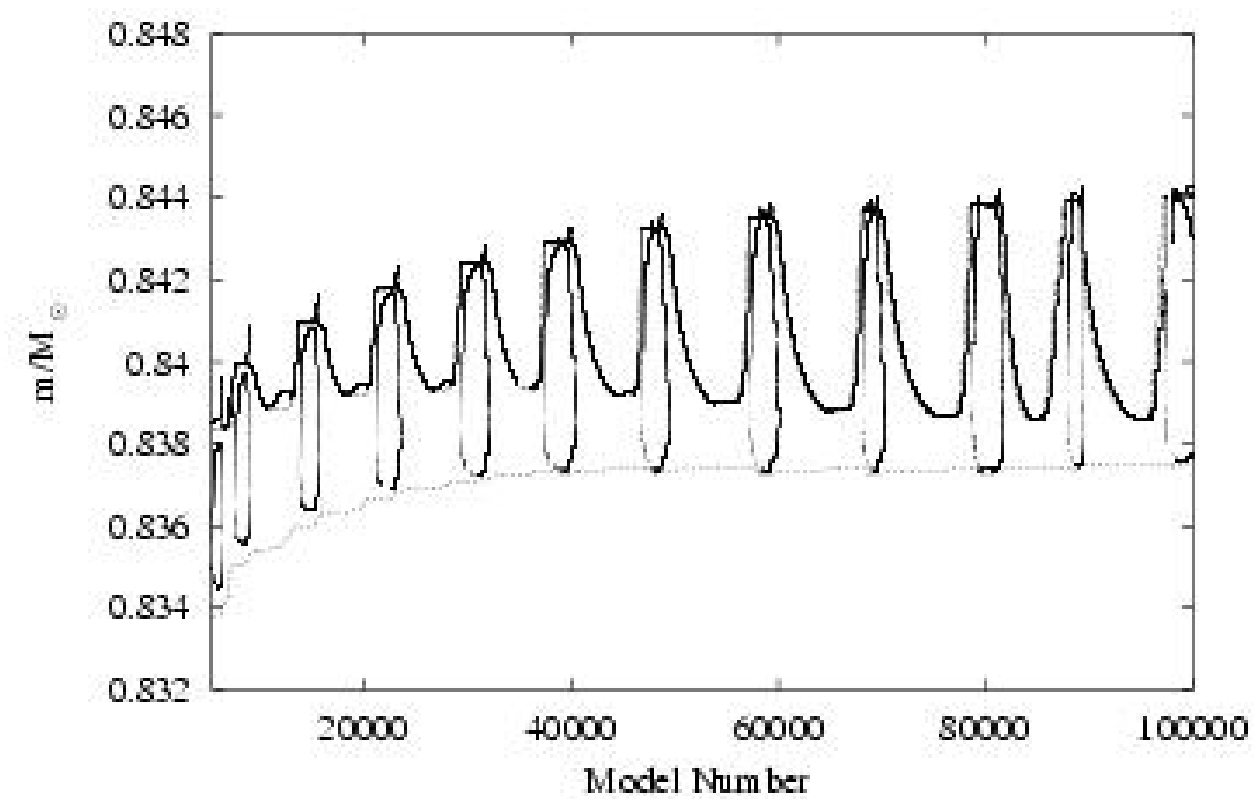}
\includegraphics[width=15cm]{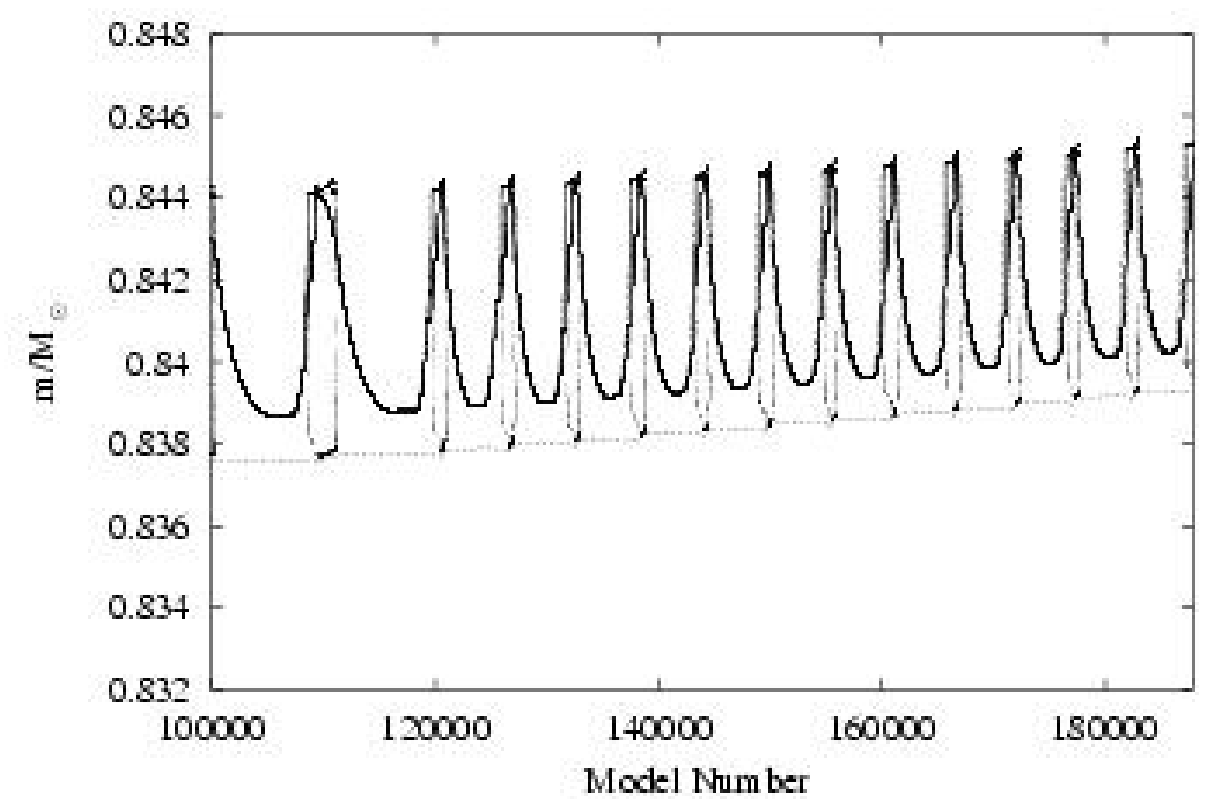}
\caption{Plot of the structure of the intershell region versus model number. The convective zone boundaries are displayed as solid lines. The hydrogen and helium burning shells are shown by short and long dashed lines respectively. Initially, efficient third dredge-up causes the intershell region to diminish. Over the latter pulses, the helium burning shell moves outward at a greater rate. We use model number rather than time in order to resolve the structure of the pulses. The ICZ typically exists for 20 years while the interpulse period is about 13,000 years.}
\label{fig:intershell}
\end{figure*}

Owing to the efficiency of third dredge-up we find that the intershell region narrows initially, as shown in Figure \ref{fig:intershell}. When the phase of most efficient dredge-up is over the intershell continues to narrow because the helium burning shell moves outward at a greater rate. This is to be expected because the helium luminosity in the interpulse phase is greater in the later pulses.

\subsection{The Effects of TDUP}

The efficiency of TDUP is usually expressed in terms of a parameter $\lambda$ which is the ratio of the amount of mass dredged up ($\Delta{M_\mathrm{DUP}}$) to the amount by which the hydrogen exhausted core grows in the preceeding interpulse ($\Delta M_\mathrm{c}$), i.e. $\lambda=\Delta{M_\mathrm{DUP}}/\Delta M_\mathrm{c}$. The value of $\lambda$ for each pulse is shown in Table \ref{tab:lambda}. We find that $\lambda$ quickly reaches a value of unity by pulse 5, in agreement with the work of \citet{Pols01}. We then find that it slightly exceeds unity for a number of pulses, before dropping back down to just below it. It should be noted that the $\lambda$ values obtained here for the first few pulses are greater than those found by \citet{Pols01} reflecting improvements in the accuracy of the solution of the equations in the core. 

\begin{table*}
\begin{center}
\begin{tabular}{c|ccccccc}
TP & M$_\mathrm{H}$/\ms & $\tau_\mathrm{ip}$/$10^4$ yrs & $\log$ (L$^{\mathrm{max}}_\mathrm{He}$/\ls) & $\Delta M_\mathrm{H}$/\ms & $\Delta M_\mathrm{DUP}$/\ms & $\lambda$ & C/O \\ 
\hline
1  & 0.83723 & ...  & 5.45274 & 0.00113 & ...     & ...   & 0.321 \\
2  & 0.83835 & 0.60 & 6.23635 & 0.00150 & 0.00097 & 0.647 & 0.321 \\ 
3  & 0.83888 & 0.69 & 6.75415 & 0.00204 & 0.00173 & 0.848 & 0.325 \\ 
4  & 0.83919 & 0.78 & 7.25165 & 0.00258 & 0.00245 & 0.950 & 0.337 \\ 
5  & 0.83932 & 0.87 & 7.68606 & 0.00308 & 0.00309 & 1.003 & 0.347 \\
6  & 0.83931 & 0.97 & 8.05930 & 0.00357 & 0.00370 & 1.036 & 0.335 \\
7  & 0.83918 & 1.07 & 8.39027 & 0.00407 & 0.00426 & 1.047 & 0.297 \\
8  & 0.83899 & 1.18 & 8.67485 & 0.00454 & 0.00473 & 1.041 & 0.237 \\
9  & 0.83880 & 1.26 & 8.88987 & 0.00493 & 0.00507 & 1.028 & 0.169 \\
10 & 0.83866 & 1.31 & 9.03736 & 0.00519 & 0.00524 & 1.012 & 0.114 \\
11 & 0.83860 & 1.34 & 9.13195 & 0.00534 & 0.00533 & 0.998 & 0.081 \\
12 & 0.83861 & 1.34 & 9.19056 & 0.00540 & 0.00531 & 0.983 & 0.067 \\
13 & 0.83870 & 1.34 & 9.22973 & 0.00542 & 0.00532 & 0.981 & 0.063 \\
14 & 0.83880 & 1.33 & 9.24354 & 0.00541 & 0.00531 & 0.981 & 0.064 \\
15 & 0.83890 & 1.31 & 9.24715 & 0.00537 & 0.00526 & 0.979 & 0.066 \\
16 & 0.83901 & 1.30 & 9.24686 & 0.00532 & 0.00521 & 0.979 & 0.068 \\
17 & 0.83912 & 1.28 & 9.24879 & 0.00528 & 0.00517 & 0.979 & 0.070 \\
18 & 0.83923 & 1.27 & 9.25030 & 0.00525 & 0.00513 & 0.977 & 0.072 \\
19 & 0.83935 & 1.26 & 9.24867 & 0.00522 & 0.00511 & 0.979 & 0.075 \\
20 & 0.83946 & 1.25 & 9.24567 & 0.00520 & 0.00508 & 0.977 & 0.078 \\
21 & 0.83958 & 1.24 & 9.24509 & 0.00517 & 0.00505 & 0.977 & 0.081 \\
22 & 0.83970 & 1.23 & 9.24255 & 0.00515 & 0.00502 & 0.975 & 0.083 \\
23 & 0.83983 & 1.22 & 9.23862 & 0.00513 & 0.00499 & 0.973 & 0.086 \\
24 & 0.83997 & 1.21 & 9.23579 & 0.00510 & 0.00497 & 0.975 & 0.089 \\
25 & 0.84010 & 1.21 & 9.23264 & 0.00508 & 0.00495 & 0.974 & 0.091 \\
\hline
\end{tabular}
\end{center}
\caption{Details of our 5\ms model. The data are: TP - the thermal pulse number, $M_\mathrm{H}$ - the hydrogen free core mass, $\tau_{\mathrm{ip}}$ - the interpulse period, L$^\mathrm{max}_\mathrm{He}$ - the peak luminosity from helium burning, $\Delta M_\mathrm{H}$ - the hydrogen free core mass growth during the interpulse, $\Delta M_\mathrm{DUP}$ - the amount of material dredged up, $\lambda$ the dredge-up efficiency parameter and C/O - the surface carbon-to-oxygen ratio by number.}
\label{tab:lambda}
\end{table*}

\begin{figure}
\includegraphics[width=8cm]{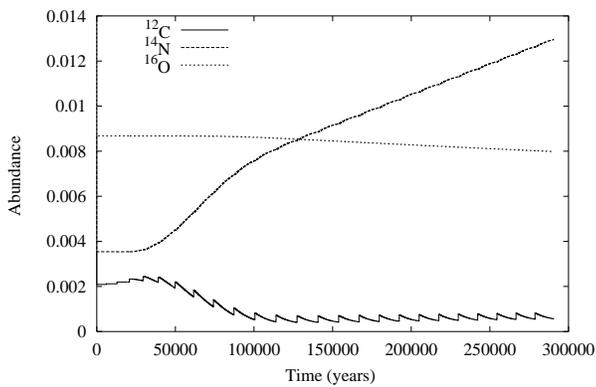}
\caption{Surface abundance evolution with time for $^{12}$C,$^{14}$N and $^{16}$O for the 5\ms\ model. While TDUP draws fresh carbon to the surface it is destroyed by hot-bottom burning during the interpulse.}
\label{fig:CNO}
\end{figure}

The evolution of the surface abundances of $^{12}$C, $^{14}$N and $^{16}$O in our model is shown in Figure \ref{fig:CNO}. Initially, the carbon abundance increases as TDUP draws carbon produced by helium burning to the surface. As dredge-up becomes deeper the envelope reaches down into hotter and hotter layers. Eventually it reaches a depth where the temperature is high enough for the CNO cycle to begin. This is hot-bottom burning (HBB). The effect of HBB is to deplete the surface of carbon and oxygen whilst enhancing the levels of nitrogen. By the $5^{\mathrm{th}}$ thermal pulse HBB is strong enough to remove more carbon than is dredged up. Hence our models show a drop in the surface carbon abundance.

Once the phase of deepest dredge-up is passed we reach a phase where carbon enhancement by TDUP is slightly greater than carbon depletion by hot-bottom burning. We find a small but persistent increase in the surface carbon abundance. The persistent rise in the nitrogen surface abundace is due to hot-bottom burning and by pulse 11 it is the most abundant of the CNO elements. The oxygen level is only slowly depeleted because the temperature of the base of the convective envelope is not high enough for the NO part of the cycle to operate efficiently.

\section{Model Comparisons}

Various groups have produced models of a 5\ms$\:$ star with Z=0.02 and no mass loss. We compare their results to ours.

The model of \citet{Stran00}, calculated using the Frascati Raphson-Newton Evolutionary Code \citep{Chieffi89}, produces a peak helium luminosity of $1.3\times10^{8}$\ls. This is lower than our value by an order of magnitude. Their model also approaches the peak luminosity much more slowly, taking around 20 thermal pulses to get there. We reach peak luminosity after just 10 pulses. As we discuss in Section 6, we suspect that the higher peak luminosity is related to the deep dredge-up we find, although we cannot verify this suspicion in this case because the dredge-up efficiency of their model is not available in the literature. Interestingly Straniero et al. do not find hot-bottom burning in their model. They give the temperature at the base of the convective envelope as $4.6\times10^7\:$K which is not hot enough for the CNO cycle to be effective. Our model has a temperature of $7.9\times10^7\:$K at the base of the envelope owing to penetration down to the deeper, hotter regions of the star.

In comparison to the model of \citet{Amand02}, produced with the Monash version of the Mount Stromlo Stellar Structure Program (MSSSP), we find a significantly smaller core mass at the first thermal pulse. Karakas et al. give a core mass of 0.862\ms whereas we calculate the value to be 0.838\ms. Comparing the efficiency of third dredge-up between the two models we find ours to be more efficient. They find $\lambda_\mathrm{max}$ of 0.955 for their equivalent model, compared to our value of 1.05. 

\section{Calculations for a 3\ms$\:$ Star}
The effects of hot-bottom burning in the 5\ms$\:$ model prevent the build-up of carbon on the surface of the star. Stars of lower mass do not have envelope temperatures high enough for HBB to occur and so there is a possibility that a carbon star could be formed. 

We apply our code to a 3\ms$\:$ star evolved from the pre-main sequence with 199 mesh points. At core helium burning the number of mesh points used is increased to 499 in order to facilitate the transition to a model with 999 mesh points when the TP-AGB is reached. We choose $\alpha=1.925$ and ignore mass loss and convective overshooting, as for the 5\ms$\:$ case.

\begin{figure}
\includegraphics[width=7cm]{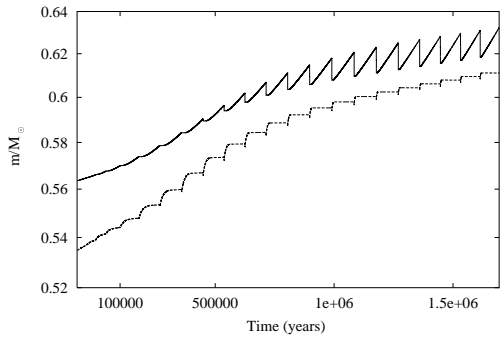}
\caption{Plot of the evolution of the hydrogen-exhausted (solid line) and helium-exhausted (dashed line) core masses with time since the peak of the first thermal pulse, for the 3\ms\ model.}
\label{fig:3core}
\end{figure}

\begin{figure}
\includegraphics[width=7cm]{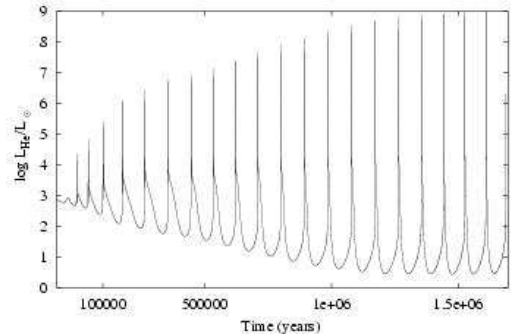}
\caption{Plot of the evolution of the luminosity from helium burning with time since the peak of the first thermal pulse, for the 3\ms\ model.}
\label{fig:LHevt}
\end{figure}

We evolve a total of 20 thermal pulse events. TDUP occurs after the 7th pulse when the hydrogen-free core mass is $0.584$\ms. A plot of the evolution of the hydrogen- and helium-free core masses is shown in Figure \ref{fig:3core}. The C/O ratio exceeds unity during TDUP following the 14th pulse at a core mass of 0.608\ms. The pulses reach a maximum helium luminosity of $\mathrm{L_{He}}=5.9\times10^8$ \ls. A plot of the evolution of the helium luminosity with time is shown in Figure \ref{fig:LHevt}.

\begin{figure}
\includegraphics[width=7cm]{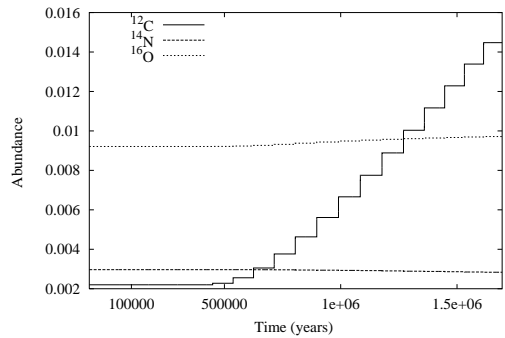}
\caption{Plot of the evolution of the CNO element abundances by mass with time since the peak of the first thermal pulse, for the 3\ms\ model.}
\label{fig:3CNO}
\end{figure}

The characteristics of the model are listed in Table \ref{tab:3lambda} and the effects of TDUP on the surface abundances by mass of the CNO elements can be seen in Figure \ref{fig:3CNO}. Initial dredge-up episodes are quite weak but the efficiency rises rapidly at each of the first few dredge-up events. Large amounts of carbon and smaller quantities of oxygen are carried to the surface. By pulse 20 dredge-up is deep with $\lambda=0.901$.

\begin{table*}
\begin{center}
\begin{tabular}{c|ccccccc}
TP & M$_\mathrm{H}$/\ms & $\tau_\mathrm{ip}$/$10^4$ yrs & $\log$ (L$^{\mathrm{max}}_\mathrm{He}$/\ls) & $\Delta M_\mathrm{H}$/\ms & $\Delta M_\mathrm{DUP}$/\ms & $\lambda$ & C/O \\ 
\hline
1  & 0.56510 & ...  & 4.36070 & 0.00108 & ...     & ...   & 0.319 \\
2  & 0.56618 & 4.29 & 4.83244 & 0.00150 & ...     & ...   & 0.319 \\
3  & 0.56768 & 5.80 & 5.41396 & 0.00233 & ...     & ...   & 0.319 \\
4  & 0.57001 & 7.77 & 6.09662 & 0.00382 & ...     & ...   & 0.319 \\
5  & 0.57383 & 8.87 & 6.42628 & 0.00480 & ...     & ...   & 0.319 \\
6  & 0.57863 & 9.17 & 6.71848 & 0.00558 & ...     & ...   & 0.319 \\
7  & 0.58421 & 9.02 & 6.88355 & 0.00617 & 0.00095 & 0.154 & 0.329 \\
8  & 0.58943 & 8.80 & 7.09441 & 0.00695 & 0.00243 & 0.350 & 0.369 \\
9  & 0.59395 & 8.78 & 7.33581 & 0.00792 & 0.00407 & 0.514 & 0.439 \\
10 & 0.59780 & 8.89 & 7.59374 & 0.00901 & 0.00587 & 0.651 & 0.539 \\
11 & 0.60094 & 9.07 & 7.84049 & 0.01014 & 0.00751 & 0.741 & 0.658 \\
12 & 0.60357 & 9.24 & 8.05703 & 0.01120 & 0.00897 & 0.801 & 0.793 \\
13 & 0.60580 & 9.34 & 8.24925 & 0.01214 & 0.01016 & 0.837 & 0.935 \\
14 & 0.60778 & 9.37 & 8.41113 & 0.01291 & 0.01114 & 0.863 & 1.084 \\
15 & 0.60955 & 9.33 & 8.54371 & 0.01351 & 0.01194 & 0.884 & 1.237 \\
16 & 0.61112 & 9.25 & 8.64986 & 0.01399 & 0.01249 & 0.893 & 1.392 \\
17 & 0.61262 & 8.97 & 8.69738 & 0.01406 & 0.01260 & 0.896 & 1.612 \\
18 & 0.61408 & 8.72 & 8.73800 & 0.01411 & 0.01270 & 0.900 & 1.695 \\
19 & 0.61549 & 8.49 & 8.77518 & 0.01414 & 0.01275 & 0.902 & 1.841 \\
20 & 0.61688 & 8.25 & 8.79855 & 0.01411 & 0.01271 & 0.901 & 1.987 \\
\hline
\end{tabular}
\end{center}
\caption{Details of our 3\ms$\:$ model. The data are: TP - the thermal pulse number, $M_\mathrm{H}$ - the hydrogen free core mass, $\tau_{\mathrm{ip}}$ - the interpulse period, L$^\mathrm{max}_\mathrm{He}$ - the peak luminosity from helium burning, $\Delta M_\mathrm{H}$ - the hydrogen free core mass growth during the interpulse, $\Delta M_\mathrm{DUP}$ - the amount of material dredged up, $\lambda$ the dredge-up efficiency parameter and C/O - the surface carbon-to-oxygen ratio by number.}
\label{tab:3lambda}
\end{table*}

\section{Detailed Model Comparisons}
Many groups have produced models with similar initial conditions to ours. Here we compare the characteristics of these models. 

The unique aspect of the {\sc{stars}} code is that it is fully self-consistent - there is no splitting into separate steps of the solution for the equations governing structure, nuclear burning and mixing of elements. Most codes solve for structure and burning together and then apply the mixing in a separate step based on this solution. This approach is not guaranteed to yield a self-consistent solution. Examples of models produced by this method include the work of \citet{Stran97} and \citet{Amand02}. Both have produced 3\ms$\:$ models at solar metallicity.

The models of \citet{Stran97} were made using the Frascati Raphson Newton Evolutionary Code \citep{Chieffi89}. They have evolved models both with and without mass loss. Details of the model without mass loss are given in Table \ref{tab:Stran3}. While the model has not been evolved to the point where it will become a carbon star, it is very likely to do so within the next few thermal pulses.

\begin{table*}
\begin{center}
\begin{tabular}{c|cccccc}
TP & M$_{\mathrm{H}}$/\ms & $\tau_{\mathrm{ip}}$/$10^4$ yrs & $\Delta M_\mathrm{H}$/\ms & $\log$ (L$^{\mathrm{max}}_\mathrm{He}$/\ls) & $\lambda$ & C/O  \\
\hline
1  & 0.572 & ...  & ...    & 5.655 & ...   & 0.316 \\
2  & 0.574 & 6.03 & 0.0016 & 4.546 & ...   & 0.316 \\
3  & 0.579 & 9.41 & 0.0054 & 6.478 & ...   & 0.316 \\
4  & 0.584 & 10.6 & 0.0050 & 6.261 & ...   & 0.316 \\
5  & 0.591 & 10.2 & 0.0062 & 6.657 & ...   & 0.316 \\
6  & 0.591 & 9.99 & 0.0062 & 6.657 & ...   & 0.316 \\
7  & 0.597 & 9.41 & 0.0066 & 6.774 & ...   & 0.316 \\
8  & 0.604 & 8.86 & 0.0067 & 6.892 & ...   & 0.316 \\
9  & 0.611 & 8.32 & 0.0069 & 6.991 & 0.029 & 0.316 \\
10 & 0.618 & 7.87 & 0.0070 & 7.086 & 0.086 & 0.322 \\
11 & 0.624 & 7.49 & 0.0072 & 7.220 & 0.153 & 0.335 \\
12 & 0.631 & 7.16 & 0.0074 & 7.322 & 0.203 & 0.357 \\
13 & 0.636 & 6.85 & 0.0074 & 7.437 & 0.257 & 0.385 \\
14 & 0.642 & 6.56 & 0.0076 & 7.538 & 0.316 & 0.420 \\
15 & 0.647 & 6.26 & 0.0077 & 7.635 & 0.338 & 0.457 \\
16 & 0.653 & 6.00 & 0.0077 & 7.716 & 0.377 & 0.497 \\
17 & 0.657 & 5.73 & 0.0077 & 7.794 & 0.390 & 0.537 \\
18 & 0.662 & 5.48 & 0.0077 & 7.832 & 0.403 & 0.578 \\
19 & 0.667 & 5.28 & 0.0076 & 7.892 & 0.434 & 0.624 \\
20 & 0.671 & 5.06 & 0.0076 & 7.961 & 0.461 & 0.672 \\
21 & 0.675 & 4.87 & 0.0076 & 8.002 & 0.461 & 0.719 \\
22 & 0.679 & 4.66 & 0.0076 & 8.046 & 0.461 & 0.797 \\
\hline
\end{tabular}
\end{center}
\caption{Details of a 3\ms$\:$ star evolved without mass-loss, taken from Table 1 of \citet{Stran97}. The data are: TP - the thermal pulse number, M$_\mathrm{H}$ - the hydrogen free core mass, $\tau_{\mathrm{ip}}$ - the interpulse period, $\Delta M_\mathrm{H}$ - the hydrogen free core mass growth during the interpulse, $\log$(L$^\mathrm{max}_\mathrm{He}$/\ls\ - the peak luminosity from helium burning, $\lambda$ the dredge-up efficiency parameter and C/O - the surface carbon-to-oxygen ratio.}
\label{tab:Stran3}
\end{table*}

Straniero et. al.'s model starts of with a slightly more massive core, $0.007$\ms$\:$ heavier than ours. They find that dredge-up does not start until the core mass has reached $0.611$\ms. In contrast we find that TDUP begins at a core mass of $0.584$\ms. Our model also shows a much deeper level of dredge-up. The peak value of $\lambda$ in Straniero et al's model is 0.461 compared with a value of 0.903 in our model. We also note a difference in the behaviour of the peak helium luminosity. The pulses of our model are more luminous and the luminosity increases more rapidly from pulse to pulse. 

The phenomena of deep dredge-up and high peak luminosities are related, as was
also found by Straniero et al. On the one hand, stronger shell flashes cause
greater expansion during the power-down phase and thereby allow the convective
envelope to reach deeper layers. Conversely, deeper dredge-up leads to a
longer interpulse period and greater values of $\Delta \mathrm{M_H}$ as found
in our model (Table~2). Hence a larger reservoir of helium is built up which
may cause a larger peak helium luminosity. Another factor that contributes to
the more powerful shell flashes is that the longer interpulse period gives the
inactive He-burning shell more time to cool and be compressed. Therefore the
next shell flash occurs at higher density and more work has to be done to
expand the intershell region, i.e. a higher peak luminosity is required.  The
upshot is that deep dredge-up and powerful shell flashes mutually reinforce
each other.

In comparision with the models of \citet{Amand02}, calculated with the Mount Stromlo Stellar Structure Program, we see similar trends. Their model has a core mass of $0.595$\ms$\:$ at the first thermal pulse and TDUP begins when the core mass reaches $0.635$\ms. These core masses are significantly heavier than those obtained in our model. They find a maximum dredge-up efficiency of $0.790$ which is also lower than the value we obtain.  

\subsection{Models with Convective Overshooting}
It is generally agreed that models produced using only convective mixing do not produce the correct abundance profiles for s-process nucleosynthesis to occur. They do not produce a pocket of $^{13}$C in the intershell region which is required to provide a neutron source via the reaction $^{13}$C($\alpha$,n)$^{16}$O. An additional mixing mechanism is required and no {\it physical} origin has yet been established. 
One of the mechanisms that have been proposed is convective overshooting, where
blobs of material penetrate into convectively stable regions leading to partial mixing of protons and carbon-rich material.

TP-AGB models with convective overshooting have been published by \citet{Her00}. He has calculated the evolution of both 3 and 4$\,$\ms\ stars in which he allows convective elements to penetrate the convectively stable regime with an exponentially decaying velocity. This overshooting is applied at all convective boundaries. These models are evolved with a Reimer's  mass-loss law
\[
\dot{M} = - 4\times10^{-13} {L R\over M} \,\mathrm{M_\odot}\, \mathrm{yr^{-1}}
\]
where $L,R$ and $M$ are the luminosity, radius and mass of the star in solar units.

Selected details of Herwig's model, taken from Table 1 of \citet{Her00}, are listed in Table \ref{tab:Her3OS}. Note that the values of $\lambda$ obtained are similar to our calculations even though we do not include overshooting in our models though our dredge-up efficiencies do not exceed unity.

\begin{table}
\begin{center}
\begin{tabular}{c|cccc}
TP & L$_\mathrm{He}$/10$^6$ \ls\ & $\lambda$ & M$_{\mathrm{H}}$/\ms\ & C/O  \\
\hline
1  &    0.46         &   -       & 0.63087 & 0.29 \\
2  &    1.17         &   -       & 0.63288 & 0.29 \\
3  &    2.13         &   -       & 0.63593 & 0.29 \\
4  &    2.63         & 0.10      & 0.63962 & 0.30 \\
5  &    3.18         & 0.23      & 0.64352 & 0.32 \\
6  &    4.63         & 0.44      & 0.64726 & 0.38 \\
7  &    6.71         & 0.53      & 0.65030 & 0.45 \\
8  &    8.80         & 0.71      & 0.65289 & 0.54 \\
9  &    13.9         & 0.82      & 0.65482 & 0.64 \\
10 &    19.0         & 0.91      & 0.65616 & 0.74 \\
11 &    29.4         & 0.98      & 0.65717 & 0.86 \\
12 &    42.9         & 1.02      & 0.65773 & 0.97 \\
13 &    59.9         & 1.04      & 0.65804 & 1.08 \\
\hline
\end{tabular}
\end{center}
\caption{Selected details of a 3\ms\ star evolved with convective overshooting and mass-loss taken from Table 1 of \citet{Her00}. The data are: TP - the thermal pulse number, L$\mathrm{He}$ - the peak luminosity from helium burning, $\lambda$ the dredge-up efficiency parameter, M$_\mathrm{H}$ - the hydrogen free core mass and C/O - the surface carbon-to-oxygen ratio.}
\label{tab:Her3OS}
\end{table}

This model naturally starts off with a higher core mass than our model because the evolution prior to the AGB with convective overshooting produces larger cores. According to Herwig the large dredge-up efficiency in his model is the result
of overshooting below the pulse-driven intershell convection zone. This leads
to a higher temperature at the bottom of this zone and a larger peak
luminosity than without overshooting. Stronger shell flashes give rise to
deeper dredge-up, as discussed above. In contrast we find similarly high peak
luminosities and similarly deep dredge-up even without overshooting.
However, it should be noted that Herwig's models produce a $^{13}$C pocket (albeit one that is too narrow - see Denissenkov \& Tout 2003) whereas ours require another mechanism. 

\section{Conclusion}
We have made a fully implicit calculation, with simultaneous solution of the structure, mixing and nuclear burning, of the evolution of both 5\ms\ and 3\ms\ stars. We have evolved the 5\ms\ star through 25 thermal pulses and their associated third dredge-up events. We find much stronger peak helium luminosities than are obtained by non-simultaneous codes and more efficient third dredge-up. Hot-bottom burning is found to occur during the interpulse. This prevents the build-up of carbon at the surface and leads to nitrogen being the most abundant CNO element.

We have followed the evolution of a 3\ms\ star through 20 thermal pulses. We find deep third dredge-up to occur after the 7$^\mathrm{th}$ thermal pulse and the star becomes a carbon star after the 14$^\mathrm{th}$ thermal pulse. We find that these events occur at much lower core masses than those found by calculations with non-simultaneous codes. We also find much more efficient third dredge-up. The presence of efficient dredge-up and low core mass may help to explain the carbon star luminosity function. Working with the results of \citet{Amand02}, Izzard \& Tout (2004) demonstrated that, in order to fit the observed LMC and SMC carbon star luminousity functions, the minimum core mass for dredge-up to occur must be $0.07$\ms\ lower than those calculated with the MSSSP. While our models are at higher metallicity, we find dredge-up to occur when the core mass is $0.584$\ms\  or $0.05$\ms\ lower than the corresponding model of \citet{Amand02}.

We conclude that our calculations yield results that differ in very important ways from calculations made with non-simultaneous codes. We find third dredge-up to be more efficient and that it occurs at lower core masses. This may be due to the simultaneous solution of the equations although there are other minor differences between the numerical and physical details of different evolution codes. It should be noted that we have not conisdered the effects of mass loss in this work. Third dredge-up depends on the envelope mass (see e.g. Straniero et al. 2003) and the inclusion of mass loss would be expect to reduce the efficiency of third dredge-up.

\section{Acknowledgements}
We thank the referee, Oscar Straniero, for his helpful comments. We thank Peter Eggleton and Rob Izzard for useful conversations. RJS would like to thank PPARC for a studentship. CAT would like to thank Churchill College for a fellowship. ORP would like to thank the Institute of Astronomy for supporting many of his visits.

\label{lastpage}

\end{document}